\documentclass{article}
\usepackage{amsmath}
\usepackage{fancyhdr}

\input{tcilatex}

\begin{document}

\title{Tradeoffs in the Quantum Search Algorithm }
\author{Lov K. Grover \thanks{%
This research was partly supported by NSA\ \&\ ARO under contract no.
DAAG55-98-C-0040.} \and \textit{lkgrover@bell-labs.com} \\
1D435 Bell Laboratories, Lucent Technologies,\\
600-700 Mountain Avenue, Murray Hill, NJ 07974}
\date{}
\maketitle

\begin{abstract}
Quantum search is a quantum mechanical technique for searching $N$
possibilities in only $\sqrt{N}$ steps. This has been proved to be the best
possible algorithm for the exhuastive search problem in the sense the number
of queries it requires cannot be reduced. However, as this paper shows, the
number of non-query operations, and thus the total number of operations, can
be reduced. The number of non-query unitary operations can be reduced by a
factor of $\log N/\alpha \log (\log N)$\ while increasing the number of
queries by a factor of only $(1+\left( \log N\right) ^{-\alpha })$. Various
choices of $\alpha $ yield different variants of the algorithm. For example,
by choosing $\alpha $ to be $O(\log N/\log (\log N))$, the number of
non-query unitary operations can be reduced by 40\% while increasing the
number of queries by just two.
\end{abstract}

\baselineskip=14pt

\newpage \pagebreak

\section{Introduction}

The quantum search algorithm was a somewhat surprising result since it gave
a means of searching $N\;$items in only $\sqrt{N}$ steps \cite{grover96}. It
was surprising because unlike most computer science applications, the
problem under consideration did not have any structure that the algorithm
could make use of. It is easy to see that any classical algorithm, whether
probabilistic or deterministic, would need $O(N)$ oracle queries - it had
generally been assumed that $O(N)$ steps would be required by \textit{any}
algorithm. However, quantum mechanical systems can be in multiple states
simultaneously and there is no clearly defined bound on how rapidly they can
search.

It was proved through subtle properties of unitary transformations that any
quantum system would need at least $O(\sqrt{N})$ queries to search $N$ items %
\cite{bbbv}. Subsequently, after the quantum search algorithm was invented,
it was proved that the number of queries required by the algorithm was
optimal and could not be improved even by one \cite{zalka}. This is usually
expressed by saying that ``the quantum search algorithm is the best possible
algorithm for exhaustive search.''

It is true that the number of queries required probably cannot be reduced,
however, as this paper shows, there is scope for improvement in the total
number of operations required by the algorithm. This is achieved by breaking
up the non-query transformations into bitwise operations in a way somewhat
reminiscent of the techniques used to improve the sorting algorithm beyond
the information theoretic limit \cite{sorting}.

It is shown that by slightly increasing the number of queries, the total
number of operations can be reduced by a logarithmic factor. This is
accomplished by making use of the amplitude amplification principle.

\section{\protect\bigskip\ Amplitude Amplification}

A few years after the invention of the quantum search algorithm, it was
generalized to a much larger class of applications known as the amplitude
amplification algorithms \cite{ampt. amp.} (similar results are
independently proved in \cite{bht}). In these algorithms, the amplitude
produced in a particular state by a unitary operation $U$, can be \textit{%
amplified} by successively repeating the sequence of operations:\ $%
Q=I_{s}U^{\dag }I_{t}U$. It was proved that if we start from the $s$ state
and repeat the operation sequence $I_{s}U^{\dag }I_{t}U,$ $\eta $ times
followed by a single repetition of $U$, then the amplitude in the $t$ state
becomes approximately 2$\eta U_{ts}$ (provided $\eta U_{ts}\ll 1$). Also, if
we start from $s$ and carry out $\frac{\pi }{4\left| U_{ts}\right| }$
repetitions of $Q$ followed by a single repetition of $U,$ we reach $t$ with
certainty. The quantum search algorithm is a particular case of this with $U$
being the Walsh-Hadamard Transformation ($W$) and $s$ being the $\overline{0}
$ state.

The power of the amplitude amplification technique lies in the fact that $%
U\; $can be \textit{any} unitary operation. Once we can design a unitary
operation (or a sequence of unitary operations)\ $U$, that produce a certain
amplitude in the target state, the amplitude amplification principle gives a
prescription for \textit{amplifying} this amplitude. The amount of
amplification increases linearly with the number of repetitions of $Q$ and
hence the probability of detecting $t$ goes up quadratically. For many
applications this results in a square-root speed up over the equivalent
classical algorithm.

In this paper we use the amplitude amplification principle for enhancing the
quantum search algorithm. This is achieved by designing a sequence of
bitwise operations that produce almost the same amplitude in the $t$ state
while requiring a much smaller number of operations.

There have been several extensions of the quantum search algorithm as well
as several applications of the algorithm to problems not immediately related
to searching; however the result presented in this paper is the first
improvement of the quantum search algorithm for the original exhaustive
search problem.

\section{The Quantum Search Algorithm}

As mentioned before, the quantum search algorithm is a particular case of
amplitude amplification with the Walsh-Hadamard Transformation being the $U$
operation and $s$ being the $\overline{0}$ state. For any $t$, $\left|
U_{ts}\right| =\frac{1}{\sqrt{N}}.$ It follows from the amplitude
amplification principle that if we start from $\overline{0}$ and carry out $%
\frac{\pi \sqrt{N}}{4}$ repetitions of the sequence of operations $-I_{%
\overline{0}}WI_{t}W,$ followed by $W$, we reach the $t$ state with
certainty. Equivalently:%
\begin{equation*}
W\underset{\frac{\pi \sqrt{N}}{4}repetitions}{\underbrace{\left( -I_{%
\overline{0}}WI_{t}W\right) \ldots \left( -I_{\overline{0}}WI_{t}W\right)
\left( -I_{\overline{0}}WI_{t}W\right) \left( -I_{\overline{0}%
}WI_{t}W\right) }}\left| \overline{0}\right\rangle =\left| t\right\rangle
\end{equation*}

Let $N$ be the number of items being searched. Then $I_{\overline{0}}$
requires us to calculate the AND\ of $\log _{2}N$ boolean variables which
can be carried out by $\log _{2}N$ C$^{2}$NOT\ operations. $W$ requires $%
\log _{2}N$ one-qubit operations. Thus the total number of\ additional
(non-query) qubit operations required by the algorithm is $\frac{\pi \sqrt{N}%
}{4}\times 3\times \log _{2}N$ while the number of queries required is $%
\frac{\pi \sqrt{N}}{4}.$ In the following section we show how to reduce the
number of additional (non-query) qubit operations while keeping the number
of queries approximately the same.

\section{Inversion about Average}

There have been several interpretations of the quantum search algorithm \cite%
{sch}. One of the ways the algorithm was first presented was in terms of an 
\textit{inversion about average} transformation \cite{grover96}. In this
paper, the inversion about average transformation is combined with the
amplitude amplification technique to obtain a faster algorithm for
exhaustive search. Before presenting the new algorithm, we first recall the
inversion about average transformation.

Consider the operation sequence: $\left( -WI_{\overline{0}}W\right) $. This
may be written as: $-W$ $(I-2\left| \overline{0}\right\rangle \left\langle 
\overline{0}\right| )W$ or equivalently $(2W\left| \overline{0}\right\rangle
\left\langle \overline{0}\right| W-I)$. The transformation $W\left| 
\overline{0}\right\rangle \left\langle \overline{0}\right| W$ can be
represented as an $N\times N$ matrix with each entry equal to $\frac{1}{N}$.
To see this recall that $W_{xy}=(-1)^{\overline{x}\cdot \overline{y}}\frac{1%
}{\sqrt{N}}$ where $\overline{x}\;\&\;\overline{y}$ denote the binary
representation of $x\;$and $y;$ $\overline{x}\cdot \overline{y}$ denotes the
bitwise dot product of $\overline{x}\;\&\;\overline{y}$. Clearly if either $%
x $ or $y$ is $0$ then $\overline{x}\cdot \overline{y}=0$ and $W_{xy}=\frac{1%
}{\sqrt{N}}.$Therefore $W\left| \overline{0}\right\rangle \left\langle 
\overline{0}\right| W$ is an $N\times N$ matrix with each entry equal to $%
\frac{1}{N}$ and each element of the transformed vector is equal to the
average of all elements of the initial vector, i.e. if the $i^{th}$
component of the input vector, $\overline{\alpha }$, is $\alpha _{i}$, then
each component of the vector $W\left| \overline{0}\right\rangle \left\langle 
\overline{0}\right| W\,\overline{\alpha }$ is $\alpha _{AV}$ where $\alpha
_{AV}\equiv \frac{1}{N}\sum_{i}\alpha _{i}$. Hence the $i^{th}$ component of
the transformed vector $(2W\left| \overline{0}\right\rangle \left\langle 
\overline{0}\right| W-I)\overline{\alpha }$ is equal to $2\alpha
_{AV}-\alpha _{i}.$ This may be written as $\alpha _{AV}-\left( \alpha
_{i}-\alpha _{AV}\right) ,$ i.e. the $i^{th}$ component in the transformed
vector is as much below the average as the $i^{th}$ component in the initial
vector was above the average, i.e. this transformation is an \textit{%
inversion about average.}

As mentioned before, the quantum search algorithm consists of the operation
sequence: $W\underset{\frac{\pi \sqrt{N}}{4}repetitions}{\underbrace{\left(
-I_{\overline{0}}WI_{t}W\right) \ldots \left( -I_{\overline{0}%
}WI_{t}W\right) \left( -I_{\overline{0}}WI_{t}W\right) \left( -I_{\overline{0%
}}WI_{t}W\right) }}\left| \overline{0}\right\rangle .$ It is insightful to
write this as: $\underset{\frac{\pi \sqrt{N}}{4}repetitions}{\underbrace{%
\left( -WI_{\overline{0}}W\right) \;I_{t}\;\ldots \;\left( -WI_{\overline{0}%
}W\right) \;I_{t}\;\left( -WI_{\overline{0}}W\right) \;I_{t}\;}}\;W\left| 
\overline{0}\right\rangle $. In terms of the inversion about average
transformations, this has the following interpretation:

\begin{enumerate}
\item $W\left| \overline{0}\right\rangle .$ The $W\;$operation applied to $%
\left| \overline{0}\right\rangle $ creates a superposition with equal
amplitude in each of the $N$ states.

\item $I_{t}$ selectively inverts the amplitude in the target state.\newline
\newline
\noindent Next the following sequence of operations (3-4) is repeated $\frac{%
\pi \sqrt{N}}{4}$ times:

\item $\left( -WI_{\overline{0}}W\right) $. As described above, this is the
inversion about average transformation. The average amplitude $\left( \alpha
_{AV}\right) $ is approximately equal to the amplitude of the $(N-1)$
non-target states. Therefore as a result of this transformation, the
amplitude in the non-target states is unaltered. Since the $t$ state is
inverted, its amplitude is below the average. As described in \cite{grover96}%
, its amplitude changes sign and its magnitude increases by 2$\alpha _{AV}.$

\item $I_{t}$ selectively inverts the amplitude in the target state thus
undoing the sign change in (3). This prepares the system for the next
inversion about average operation through which the magnitude of the
amplitude in the $t$ state is increased.
\end{enumerate}

\bigskip \bigskip 

\textbf{Figure 1 (see attached file): The transformation }$-\left( W^{(S)}I_{%
\overline{0}}^{(S)}W^{(S)}\right) $\textbf{\ performs an inversion about
average in each of the four subsets of states - the four subsets are defined
by the condition that the qubits not in S\ stay fixed (in the above figure,
the qubits not in S\ are qubits 1 \& 2), e.g. in the first subset, qubits 1
\&\ 2 are both 0.}

\section{Partial Inversion About Average}

Assume there to be $n$ qubits, then as described in the previous section, $%
\left( -WI_{\overline{0}}W\right) $ does an inversion about average
transformation on the entire set of $N\equiv 2^{n}$ states. Consider a set
that contains $m$ of the $n$ qubits, denote this set by $S$. Define the
Walsh-Hadamard transformation on $S$ as the operation $H=\frac{1}{\sqrt{2}}%
\left[ 
\begin{array}{cc}
1 & 1 \\ 
1 & -1%
\end{array}%
\right] $, applied to each qubit in the set $S$ and denote this by $W^{(S)}$%
. Similarly define the operation $I_{\overline{0}}^{(S)}$ as the selective
inversion of the state in which each qubit in $S$ is 0.

Consider the following transformation $-\left( W^{(S)}I_{\overline{0}%
}^{(S)}W^{(S)}\right) .$ Its effect is to partition the states into subsets
such that in each subset the qubits that are not in $S$ stay fixed. This
transformation leaves the total probability in each subset the same - within
each subset, an \textit{inversion about average} transformation takes place.
In the above figure the set $S$\ contains qubits 3 \&\ 4. It partitions the
state into 4 subsets in which the qubits \textit{not} in the set are fixed,
e.g. in the first subset, qubits 1 \& 2 are both 0. The transformation $%
-\left( W^{(S)}I_{\overline{0}}^{(S)}W^{(S)}\right) ,$ does an inversion
about average separately in each of the four subsets.

\section{Improved Quantum Search Algorithm}

As discussed above, the quantum search algorithm increases the amplitude in
the $t$ state through successive repetitions of selective inversion and
inversion about average. The inversion about average operation increases the
amplitude in the $t$ state by an amount equal to the average amplitude over
all states. The inversion about average requires three transformations: $%
W,\;I_{\overline{0}}\;\&\;W$ each of which requires $\log _{2}N$ qubit
operations. In the following, we show how to carry out the inversion about
average transformations over a smaller subset of states thus requiring fewer
than $\log _{2}N$ qubit operations.

\subsection{Basic $U$ operation}

As mentioned earlier in section 3, the amplitude amplification principle
requires a basic transformation $U$ that produces a certain transition
amplitude, $U_{ts}$ from $s$ to $t$. This can then be iterated as in section
3, to amplify the amplitude in $t$.

Divide the $\log _{2}N$ qubits used to represent the $N$ items into sets of $%
\alpha \log _{2}(\log _{2}N)$ qubits ($\alpha >1)$. Since there are $\log
_{2}N$ qubits, there will be $\eta \equiv \frac{\log _{2}N}{\alpha \log
_{2}(\log _{2}N)}$ sets. Define the Walsh-Hadamard transformation on the $%
i^{th}$ set as the operation $H=\frac{1}{\sqrt{2}}\left[ 
\begin{array}{cc}
1 & 1 \\ 
1 & -1%
\end{array}%
\right] $, applied to each qubit in the set and denote this by $W^{(i)}$.
Similarly define the operation $I_{\overline{0}}^{(i)}$ as the selective
inversion of the state in which each qubit in the $i^{th}$ set is 0.

Next consider the following transformation:%
\begin{equation*}
U\equiv \left( -W^{(\eta )}I_{\overline{0}}^{(\eta )}W^{(\eta )}\right)
I_{t}\ldots \left( -W^{(i)}I_{\overline{0}}^{(i)}W^{(i)}\right) I_{t}\ldots
\left( -W^{(1)}I_{\overline{0}}^{(1)}W^{(1)}\right) I_{t}\;\;W
\end{equation*}

\bigskip \bigskip 

\textbf{Figure 2 - (see attached file) - The inversion about average
transformation in the standard quantum search algorithm \smallskip\ is
replaced by two such operations - one that acts on the horizontal sets \&\
the other on the vetical sets.}

\medskip 

When applied to the $\left| \overline{0}\right\rangle $ state, it has the
following effect:

\begin{enumerate}
\item The first application of $W$ produces a superposition with equal
amplitudes in all states.

\noindent \noindent After this, each application of $\left( -W^{(i)}I_{%
\overline{0}}^{(i)}W^{(i)}\right) I_{t}$ does the following.

\item $I_{t}$ inverts the amplitude in the target state.

\item $\left( -W^{(i)}I_{\overline{0}}^{(i)}W^{(i)}\right) $ does a partial 
\textit{inversion about average} in each subset of states defined by the
condition that the state of all qubits \emph{not} in the $i^{th}$ set stays
constant (as shown in figure 1).
\end{enumerate}

Next consider the effect of steps 2 \& 3 on the subset of states that
contains $t$. Let the amplitude of the $t$ state be $\frac{a}{\sqrt{N}}$.
After step 2, the amplitude of $t$ becomes $-\frac{a}{\sqrt{N}}.$

After step 2, the amplitude of each of the other states in the subset
containing $t$ is the same as after step 1, i.e. $\frac{1}{\sqrt{N}}$. This
is because the first $(i-1)$ inversion about average transformations act on
subsets of states in which the value of the $i^{th}$ qubit is constant.
Hence they produce no change in the amplitude of any state in which the
value of the $i^{th}$ qubit is different from the value of the $i^{th}$
qubit in the $t$ state.

The number of states in each subset is $2^{\alpha \log _{2}(\log _{2}N)}$
which is $\left( \log _{2}N\right) ^{\alpha }$. Therefore the average
amplitude in the $i^{th}$ subset of states containing $t$ is $\frac{1}{\sqrt{%
N}}-\frac{a+1}{\left( \log _{2}N\right) ^{\alpha }\sqrt{N}}$. Step 3 (the
partial inversion about average), increases the amplitude in $t$ to $\frac{a%
}{\sqrt{N}}+2\left( \frac{1}{\sqrt{N}}-\frac{a+1}{\left( \log _{2}N\right)
^{\alpha }\sqrt{N}}\right) $.

Assuming $a<\log _{2}N,$ the increase in amplitude of $t$ due to 2 \&\ 3 is
at least $2\left( \frac{1}{\sqrt{N}}-\frac{1}{\left( \log _{2}N\right)
^{\alpha -1}\sqrt{N}}\right) .$ Therefore in the $\eta $ repetitions of 2 \&
3, the amplitude of $t$ increases by at least $2\eta \left( \frac{1}{\sqrt{N}%
}-\frac{1}{\left( \log _{2}N\right) ^{\alpha -1}\sqrt{N}}\right) .$

The operation $U$ described by 1, 2 \& 3 above, forms the building block for
the amplitude amplification algorithm described in the following section.

\subsection{Amplitude Amplification\newline
}

As described in the analysis above, the composite operation $U$ when applied
to $\left| \overline{0}\right\rangle $ produces an amplitude of at least $%
\frac{1}{\sqrt{N}}\left( 2\eta \left( 1-\frac{1}{\left( \log _{2}N\right)
^{\alpha -1}}\right) +1\right) $ in $t$. Therefore by the amplitude
amplification principle, $\frac{\pi \sqrt{N}}{4}\frac{1}{2\eta \left( 1-%
\frac{1}{\left( \log _{2}N\right) ^{\alpha -1}}\right) +1}$ \ repetitions of
the $I_{s}U^{^{\dag }}I_{t}U$ operation sequence followed by a single
application of $U$, \ will concentrate the amplitude in the $t$ state.

Note that $U^{^{\dag }}$ consists of the same operations as $U$ but in the
opposite order:%
\begin{equation*}
U^{^{\dag }}\equiv W\;I_{t}\left( -W^{(1)}I_{\overline{0}}^{(1)}W^{(1)}%
\right) \;\ldots \;\;I_{t}\left( -W^{(i)}I_{\overline{0}}^{(i)}W^{(i)}%
\right) \;\ldots \;I_{t}\left( -W^{(\eta )}I_{\overline{0}}^{(\eta
)}W^{(\eta )}\right)
\end{equation*}

\subsection{Analysis}

Each application of $U$ requires $\eta $ queries. Therefore in each
application of $I_{s}U^{^{\dag }}I_{t}U$ there are $\left( 2\eta +1\right) $
queries. Neglecting the single application of $U$ at the end, it follows
that the total number of queries is: 
\begin{equation*}
\left( 2\eta +1\right) \times \frac{\pi \sqrt{N}}{4}\frac{1}{2\eta \left( 1-%
\frac{1}{\left( \log _{2}N\right) ^{\alpha -1}}\right) +1}.
\end{equation*}

\noindent This is less than $\frac{\pi \sqrt{N}}{4}\frac{1}{\left( 1-\frac{1%
}{\left( \log _{2}N\right) ^{\alpha -1}}\right) }$.

The total number of applications of $U$ in the algorithm is $2\times \frac{%
\pi \sqrt{N}}{4}\frac{1}{2\eta \left( 1-\frac{1}{\left( \log _{2}N\right)
^{\alpha -1}}\right) +1}$ (as before, neglecting the single application of $%
U $ at the end). The number of additional (non-query) qubit operations
required in each application of $U$ is log$_{2}N+3\times \eta \times \alpha
\log _{2}(\log _{2}N)$ which is equal to $4\log _{2}N$. The total number of
additional (non-query) qubit operations due to the $U$ \& $U^{^{\dag }}$
hence becomes $\frac{2\pi \sqrt{N}\log _{2}N}{2\eta \left( 1-\frac{1}{\left(
\log _{2}N\right) ^{\alpha -1}}\right) +1}.$ In addition there are $\frac{%
\pi \sqrt{N}}{4}\frac{1}{2\eta \left( 1-\frac{1}{\left( \log _{2}N\right)
^{\alpha -1}}\right) +1}$ $I_{s}$ operations each of which requires $\log
_{2}N$ operations. Therefore the total number of additional (non-query)
qubit operations required is $\frac{2\pi \sqrt{N}\log _{2}N}{2\eta \left( 1-%
\frac{1}{\left( \log _{2}N\right) ^{\alpha -1}}\right) +1}\times \frac{9}{8}$%
. This is less than $\frac{9}{8}\pi \alpha \sqrt{N}\log _{2}\left( \log
_{2}N\right) $ provided $\alpha \geq 2.$

\section{Comparison}

The quantum search algorithm needs $\frac{\pi \sqrt{N}}{4}$ queries and $%
\frac{3\pi \sqrt{N}\log _{2}N}{4}$ additional (non-query) qubit operations.

The algorithm of the previous section needs fewer than $\frac{\pi \sqrt{N}}{4%
}\frac{1}{\left( 1-\frac{1}{\left( \log _{2}N\right) ^{\alpha -1}}\right) }$
queries and less than $\frac{9}{8}\pi \alpha \sqrt{N}\log _{2}\left( \log
_{2}N\right) =\frac{9\pi \sqrt{N}\log _{2}N}{8\eta }$ additional (non-query)
qubit operations (provided $\alpha \geq 2)$, note that the ratio of the
additional (non-query)\ qubit operations required by the two algorithms is $%
\frac{3}{2\eta }$

\subsection{Smallest increase in the number of queries}

In case ($\alpha -1)$ is $\frac{\log _{2}N}{2\log _{2}\left( \log
_{2}N\right) }$, then the number of queries required by the improved
algorithm is less than $\frac{\pi \sqrt{N}}{4}\frac{1}{\left( 1-\frac{1}{%
\sqrt{N}}\right) }$, i.e. the increase in the number of queries as compared
to that required by the standard quantum search algorithm seems to be less
than one. However this is only suggestive since several other effects become
significant when $\alpha $ becomes this large (and therefore $\eta ,$ the
number of sets of qubits, which was$\frac{\log _{2}N}{\alpha \log _{2}(\log
_{2}N)},$ becomes small). In fact the smallest value for $\eta $ is 2. We
analyze this case separately below.

This is perhaps the simplest example of the partial inversion about average.
The qubits are partitioned into two sets with $\frac{1}{2}\log N$ qubits in
each set. Then the basic $U$\ operation is the following:

\begin{equation*}
U\equiv \left( -W^{(2)}I_{\overline{0}}^{(2)}W^{(2)}\right) I_{t}\left(
-W^{(1)}I_{\overline{0}}^{(1)}W^{(1)}\right) I_{t}\;\;W
\end{equation*}

A simple analysis shows that the amplitude in the $t$ state after applying $%
U $ to the $0$ state (which is $U_{ts}$) becomes $\frac{5}{\sqrt{N}}-\frac{12%
}{N}+O\left( \frac{1}{N^{1.5}}\right) .$An amplitude amplification as
described previously in this paper will now amplify this amplitude.

To compare this to the standard quantum search algorithm, observe that the
standard quantum search algorithm is obtained by taking $U$ to be as follows:

\begin{equation*}
U\equiv \left( -WI_{\overline{0}}W\right) I_{t}\left( -WI_{\overline{0}%
}W\right) I_{t}\;\;W
\end{equation*}

This produces a $U_{ts}$ of $\frac{5}{\sqrt{N}}+O\left( \frac{1}{N^{1.5}}%
\right) $. Since the number of queries is known to be proportional to $%
U_{ts} $, the number of additional queries required by the new algorithm is
obtained by scaling the queries required by the standard quantum search.
This gives the number of additional queries as approximately: $\frac{\pi 
\sqrt{N}}{4}\times \frac{12}{5\sqrt{N}}\simeq 2.$ Note that such a small
increase in the number of queries is not likely to be significant since it
would typically take the quantum search algorithm $\frac{\pi \sqrt{N}}{4}\pm
O(1)$ queries to go from an approximate to the exact solution.

The number of additional (non-query) qubit operations required can be
compared to the standard quantum search by comparing the two $U$ operations.
Assuming each $W$ and $I_{\overline{0}}$ need twice the number of operations
as compared to $W^{(1)}$, $W^{(2)},I_{\overline{0}}^{(1)},I_{\overline{0}%
}^{(2)}$, it follows that the new algorithm will need only $\frac{3}{5}$ as
many operations as compared to standard quantum searching.

\subsection{Minimizing the total number of operations}

If we permit a very slight increase in the number of queries, the number of
additional unitary operations and hence the total number of operations can
be significantly reduced.

Assume that each query requires $K\log _{2}N$ qubit operations, where $K$ is
order 1. This is plausible since the query is a function of $\log _{2}N$
qubits and so would need $O(\log _{2}N)$ steps to evaluate. The total number
of qubit operations is hence approximately:%
\begin{eqnarray*}
&&K\log _{2}N\times \frac{\pi \sqrt{N}}{4}\frac{1}{\left( 1-\frac{1}{\left(
\log _{2}N\right) ^{\alpha -1}}\right) }+\frac{9}{8}\pi \alpha \sqrt{N}\log
_{2}\left( \log _{2}N\right) \\
&\approx &K\log _{2}N\times \frac{\pi \sqrt{N}}{4}\left( 1+\frac{1}{\left(
\log _{2}N\right) ^{\alpha -1}}\right) +\frac{9}{8}\pi \alpha \sqrt{N}\log
_{2}\left( \log _{2}N\right)
\end{eqnarray*}

\noindent Differentiating with respect to $\alpha $ and setting the
derivative to zero gives the condition:%
\begin{equation*}
-K\log _{2}N\times \frac{\pi \sqrt{N}}{4}\frac{\log _{e}\left( \log
_{2}N\right) }{\left( \log _{2}N\right) ^{\alpha -1}}+\pi \sqrt{N}\log
_{2}\left( \log _{2}N\right) =0
\end{equation*}

\noindent This gives $\left( \log _{2}N\right) ^{\alpha -2}=\frac{K\log _{e}2%
}{4}.$ Substituting in the expression for the total number of operations
gives:%
\begin{eqnarray*}
&&\frac{\pi \sqrt{N}}{4}K\log _{2}N+\frac{\pi \sqrt{N}}{\log _{e}2}+\frac{9}{%
4}\pi \sqrt{N}\log _{2}\left( \log _{2}N\right) +\pi \sqrt{N}\log _{2}\frac{%
K\log _{e}2}{4} \\
&\approx &\frac{\pi \sqrt{N}}{4}K\log _{2}N+\frac{9}{4}\pi \sqrt{N}\log
_{2}\left( \log _{2}N\right)
\end{eqnarray*}

In comparison the standard quantum search algorithm requires$\frac{\pi \sqrt{%
N}}{4}K\log _{2}N+\frac{3\pi \sqrt{N}\log _{2}N}{4}$ qubit operations.
Therefore the number of additional two-qubit operations has been reduced by
a factor of $\frac{\log _{2}N}{3\log _{2}\left( \log _{2}N\right) }.$

\section{Further Improvements?}

The goal of this paper is to make a statement that the quantum search
algorithm can be further improved. It is hoped that this will lead to
further research in this direction. There is scope for further improvements
in the algorithm presented in this paper, though at the cost of more
complicated calculations. Some of these improvements, such as modifying the $%
U$ operation to include multiple inversions about average in each subset,
are being explored and will be presented in more detail later.

\end{document}